# MEG Evidence That Modality-Independent Conceptual Representations Encode Visual but Not Lexical Representations

Julien Dirani [*1] & Liina Pylkkänen[1,2,3]

[1]Psychology Department, New York University, New York, NY, USA, 10003
[2]Linguistics Department, New York University, New York, NY, USA, 10003
[3]NYUAD Institute, New York University Abu Dhabi, Abu Dhabi, UAE, 129188

**Abstract**
The semantic knowledge stored in our brains can be accessed from different stimulus modalities. For example, a picture of a cat and the word "cat" both engage similar conceptual representations. While existing research has found evidence for modality-independent representations, their content remains unknown. Modality-independent representations could be abstract, or they might be perceptual or even lexical in nature. We used a novel approach combining word/picture cross-condition decoding with neural network classifiers that learned latent modality-independent representations from MEG data. We then compared these representations to models representing semantic, sensory, and lexical features. Results show that modality-independent representations are not strictly amodal; rather, they also contain visual representations. There was no evidence that lexical properties contributed to the representation of modality-independent concepts. These findings support the notion that perceptual processes play a fundamental role in encoding modality-independent conceptual representations. Conversely, lexical representations did not appear to partake in modality-independent semantic knowledge.

**INTRODUCTION**

How does the brain represent semantic knowledge? A fundamental aspect of the brain's capacity to encode meaning lies in its ability to abstract away from an initial input modality. For example, reading the word "cat" or looking at the picture of a cat both activate similar conceptual content. While a large body of work has established that different tasks and stimulus modalities (such as pictures and words) activate common brain areas as well as shared representations of concepts [1-6], the content of those representations remains unknown. A major challenge lies in the difficulty to distinguish between the processes by which semantic information is retrieved and the actual content of the representations [7]. In this paper, we define modality-independent representations as any representations that are consistently activated across multiple stimulus modalities. We use a novel representational learning approach to test the extent to which modality-independent representations contain amodal semantic features, sensory components, or properties of the lexical items corresponding to the concept. Further, we assess the extent to which the content of modality-independent representations dynamically changes over the milliseconds following the perception of a semantically engaging stimulus.

Prominent theories of word meaning and concepts, like the prototype and exemplar theories [8, 9], rely on semantic features as their foundational element of representation (e.g. is an animal, can fly, has feathers). Semantic feature representations are also employed in numerous models of semantic memory, object recognition, and word recognition [10-14]. Given the well-established predictive power of semantic feature models across multiple modalities, the most intuitive hypothesis is that they are the content of modality-independent representations. Alternatively, modality-independent representations could also contain sensory-level representations, such as visual or auditory representations. For example, reading the word "cat" could activate visual representations related to cats. In fact, several studies have shown that the brain engages sensory-motor features in the representation of concepts, such as visual shapes, sounds, and motor representations [15-17]. However, it remains unknown whether these representations are modality-independent, that is, activated for all input modalities. Concepts could also be inherently connected to language, in a way that the lexical labels of concepts are automatically activated for all stimulus modalities. For example, looking at a picture of a cat could automatically activate the lexical label "cat".

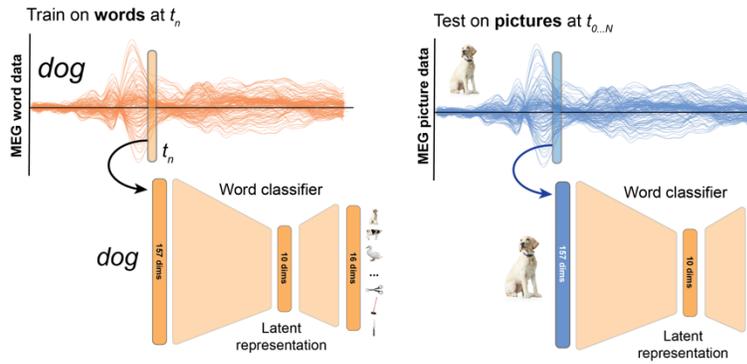

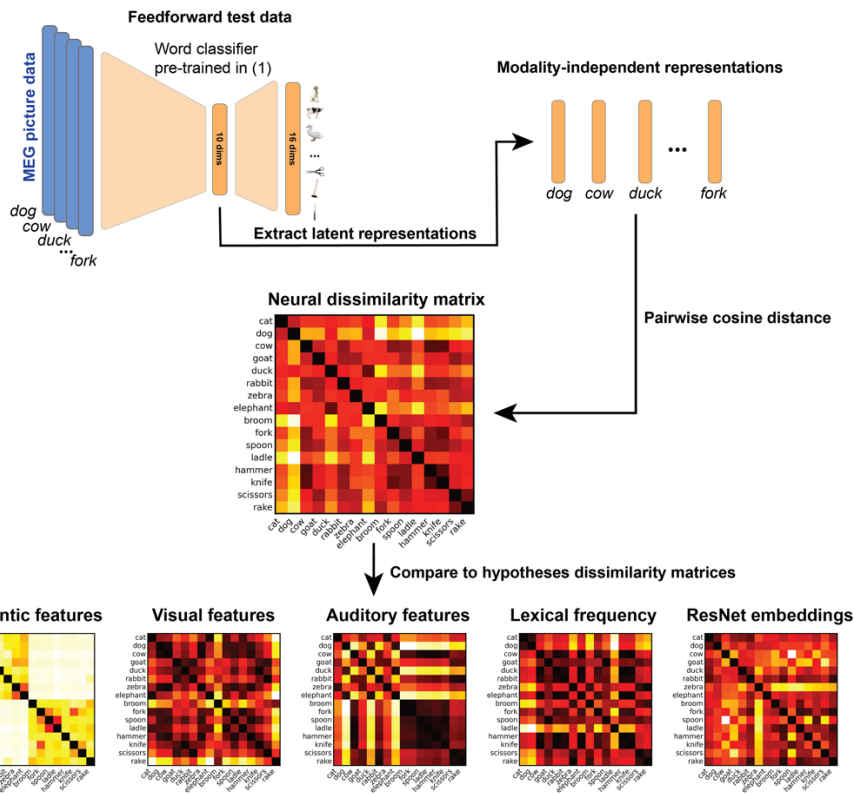

**Fig. 1: Analysis pipeline.** (1) Cross-condition decoding in which neural network classifiers trained on one modality were tested on the other modality, for all pairs of timepoints. This allowed us to map the clusters of timepoints ($t_{train}$, $t_{test}$) where modality-independent representations of basic-level concepts were activated. (2) Classifiers with successful cross-condition generalization were assumed to have learned latent representations of the semantic space that were modality-independent. We investigated the representational content of those representations using representational similarity analysis (RSA) and compared them to 4 hypotheses spaces (semantic, visual, auditory, and lexical representations). We also compared them to the ResNet embeddings of the picture stimuli to test whether the shared representations between pictures and words did not merely result from an automatic reactivation of stimulus-specific visual representations.

To address these hypotheses, we combined a cross-condition decoding approach [18] with representational similarity analysis (RSA) [19] using MEG (Fig. 1). 25 participants viewed pictures and words while making a binary animacy judgment. We used deep neural network classifiers which learned latent representations during training. If these classifiers trained on one modality (e.g., words) successfully generalize to another modality (pictures), they would have learned latent representations of concepts that are independent of the training and testing modalities. The RSA allowed us to compare these latent modality-independent representations with models representing conceptual, sensory, and linguistic aspects of the concepts.

An experimental paradigm featuring repeated presentations of pictures and their corresponding words opens up the possibility that encountering a concept in one modality could trigger visual imagery of the stimulus expressing that concept in the other modality, that is, the word 'dog' could trigger mental imagery of a previously presented picture of a dog. To address this type of interpretation for any potential modality-independent representation, we also compared the latent modality-independent representations to the stimulus-specific visual representations captured by a pretrained ResNet convolutional neural network [20].

**RESULTS**

**Modality-independent representations of basic-level concepts are activated at ~250ms**

For each timepoint $t_{train}$ of the MEG data of the word modality, we trained a classifier to decode the exemplar that was presented on screen. We then tested the classifier at all timepoints $t_{test}$ of the MEG data of the picture modality. This allowed us to find the pairs of timepoints ($t_{train}$, $t_{test}$) where modality-independent representations of concepts are activated. The results showed that for both pictures and words, modality-independent representations were active at around 250ms and sustained until around 600ms after the onset of the stimulus. Some representations that were active early in the pictures (~100-300ms) seemed to be also activated later in the words (~400-600ms) as depicted by the off-diagonal cluster on the matrix of scores (Fig. 2C) ($p$ <0.05). Notably, the converse approach of training on the picture data and testing on the words did not yield above-chance decoding accuracies suggesting that models trained on pictures may have relied on picture-specific representations that are not activated when processing words. This discrepancy likely stems from the fact that MEG data from pictures potentially carries a stronger signal reflecting low-level visual features, which does not readily generalize to words. When classifiers are trained on words, they might more effectively pick up on semantic representations since little to no information about the exemplars is present in the visual features. These classifiers would then more effectively generalize to the picture recognition process.

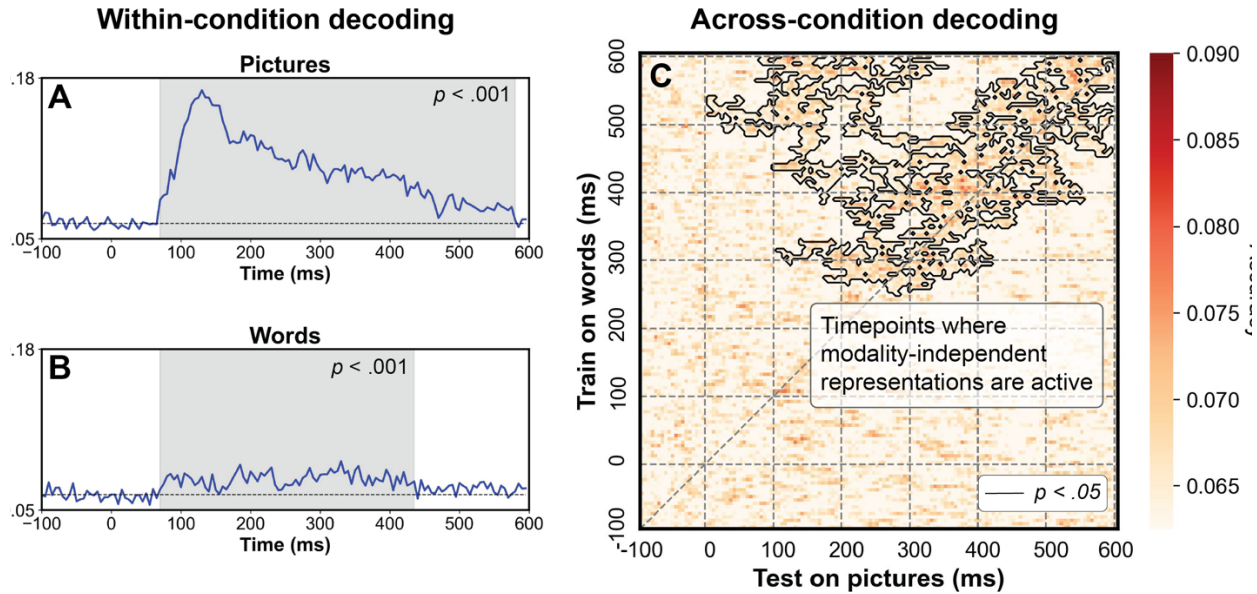

**Fig 2. The activation timing of modality-independent representations of basic-level concepts. A-B**: Accuracy scores at each time-point for classifiers trained and tested within each modality. The shaded regions indicate time points where classifier accuracy was above chance at the group level. **C**: Cross-condition decoding results where models trained on the MEG data from the words were tested on MEG data from the pictures for all pairs of timepoints ($t_{word}, t_{picture}$). The contour plot indicates the cluster of time-points with accuracy scores significantly above chance. Modality-independent representations were active at around 250ms and sustained until around 600ms after stimulus onset. The part of the cluster that is off-diagonal indicates that representations that were active earlier in the pictures (~100-300ms) were delayed in the words (~400-600ms). Models trained on MEG data from the pictures and tested on the words did not significantly surpass chance-level accuracy.

**Modality-independent conceptual representations encode semantic features and visual representations**

To assess the content of modality-independent representations, we compared the latent representation space learned by successful modality-independent classifiers to our hypotheses using RSA. This was done at each pair of timepoints ($t_{word}$, $t_{picture}$) where modality-independent representations were identified. Each hypothesis was formulated as a representational dissimilarity matrix (RDM), representing the structure of the stimulus space through pairwise distances of all exemplars. The semantic feature hypothesis was operationalized using human normed semantic features [21], while visual and auditory sensory hypotheses were operationalized using brain-based experiential norms [17]. For the lexical hypothesis, lexical frequency measures were taken from the English Lexicon Project [22].

The results showed that modality-independent representations significantly correlated both with the semantic features and the visual features (Fig. 3A-B). There was no evidence that modality-independent representations correlated with lexical representations or with auditory features (Fig. 3C-D).

The semantic feature hypothesis correlated with modality-independent representations at virtually all timepoints ($t_{word}$, $t_{picture}$) where modality-independent representations were observed, illustrating the widespread presence of semantic feature representations in the encoding of modality-independent conceptual representations. This contrasts with the visual representations which were constrained to a cluster falling around the diagonal of the accuracy matrix (Fig. 3B), indicating that the modality-independent representations that correlated with visual features evolved at relatively concurrent timepoints in the pictures and words from around 250ms until 600ms. To test whether our positive results for the presence visual features in modality independent representations may simply reflect reactivation of a specific previously encountered picture in the experiment, we processed the picture stimuli through a pre-trained convolutional neural network (ResNet) [20] and extracted the activation of the last hidden layer for each unique exemplar. This revealed no evidence that the modality-independent representations identified by the classifier reflected reactivation of specific picture stimuli when reading their corresponding words during the experimental task (Fig. 3E).

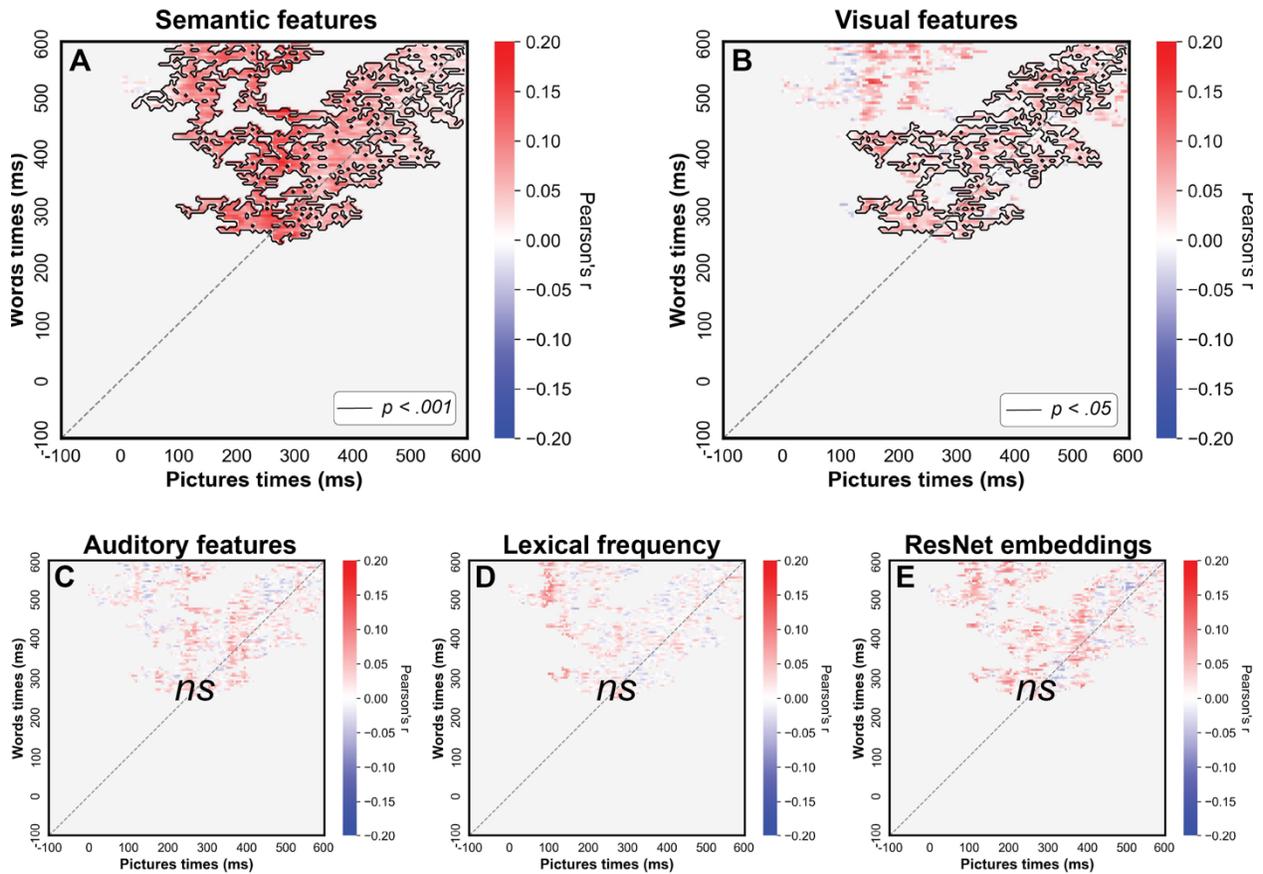

**Fig 3. RSA results investigating the content of modality-independent representations.** For each pair of timepoints ($t_{word}$, $t_{picture}$) where modality-independent representations were identified, we investigated their content using representational similarity analysis (RSA). The contour plots indicate clusters of time-points where modality-independent representations significantly correlated with the hypothesis. **A-B:** Modality-independent representations significantly correlated with the semantic features and the visual features hypothesis. Semantic features had a widespread correlation over most of the cluster, while visual representations were constrained to the part of the cluster falling around the diagonal. **C-D:** Modality-independent representations did not correlate with auditory features or lexical frequency. **E:** Modality-independent representations did not correlate with the ResNet embeddings of the picture stimuli, suggesting that shared representations between pictures and words did not merely result from an automatic reactivation of stimulus-specific visual representations.

## DISCUSSION

**Modality-independent conceptual representations are not strictly amodal**

While previous work has shown evidence that the brain stores modality-independent representations of concepts [16], to date, the content of those representations has remained unknown. Here we used an innovative approach combining cross-condition decoding with RSA to directly investigate the nature of modality independent representations. We found that modality-independent representations most strongly align with the semantic feature hypothesis, suggesting a shared semantic space across modalities. In line with expectations, the brain recruits a semantic feature space that serves to define the content of the concept irrespective of the initial stimulus modality (e.g. a dog: it is an animal, it barks, it has a tail). But this does not mean that modality-independent representations must be purely amodal, that is, void of any sensory-motor content. Semantic features often themselves constitute complex concepts (such as 'has wings'), which means they may themselves have an internal structure containing say, visual or motor representations. Consistent with this, our RSA results indicated that modality-independent representations shared a similar structure to the visual hypothesis. This suggests that the semantic content that is consistently activated for both words and pictures contains visual representations. For example, both reading the word "cat" or seeing a cat would activate some visual representations related to cats, as expected under theories in which concepts are neurally represented at least to some extent using the perceptual systems in which they are experienced [15, 17, 23-25]. In contrast, we found no evidence that modality-independent representations contain auditory representations. Of course, both our words and pictures were presented visually. Were the words presented auditorily, or the concepts as auditory objects (such as a dog barking), the recruited sensory representations might have been more auditory in nature. This can be straightforwardly tested with our novel method in the future.

**No evidence for lexical content in modality-independent representations**

Human use language to express conceptual knowledge. Thus, it is plausible that conceptually contentful stimuli could activate lexical representations even if the stimuli are not linguistic. Under this hypothesis, one would expect a correlation between modality-independent representations and lexical representations. We found no evidence for such a correlation, suggesting that lexical labels themselves are not necessarily activated when recognizing conceptual content via a non-linguistic stimulus.

**The timing of activation of modality-independent representations**

Here we trained classifiers of basic-level concepts on MEG data from the words and tested these classifiers on the MEG data from the pictures. By systematically examining all pairs of timepoints ($t_{word}$, $t_{picture}$) we mapped out the time-course of activation of modality-independent conceptual representations [18]. Our findings revealed that modality-independent representations of basic-level concepts are activated at around 250ms for both the pictures and the words. This novel finding complements prior MEG evidence that modality-independent representations of higher-level categories (animacy) are simultaneously activated at 150ms for picture and words [1]. The temporal pattern aligns with prior research in object recognition, showing that superordinate category representations are active within the first 150ms [26-28] followed by basic level representations at around 200ms [29-34]. Crucially, our results extend this temporal pattern to the activation time-course of modality-independent representations indicating that they follow a hierarchical process in which superordinate representations are activated first, followed by the basic-level representations.

**Limitations**

Here we defined modality-independent representations as any representations that are coactivated in picture and word recognition, with the assumption that these shared representations reflect a core part of the conceptual representations. However, they could also be a result of spreading activation from one modality-specific representation (e.g., lexical) to another modality-specific representation (e.g., visual), with neither corresponding to shared modality-independent representations. Disentangling these two possibilities has long been a challenge in the field, and drawing a clear distinction between then remains a challenge [35-37]. Nevertheless, here we propose a novel method that, at the very least, provides insight into the content of those representations that are systematically activated across pictures and words, indicating that they engage perceptual processes. Another limitation concerns normative models of semantic features [21] in which the features constituting a concept can themselves be complex concepts (e.g., 'has wings'), potentially incorporating other representations such as nested semantic features (e.g., 'has feathers') or perceptual representations [17]. Despite ensuring that each hypothesis contributed to a unique aspect of the representation space (with all pairwise correlations between hypothesis models < 0.3), this limitation could explain the overlap observed in the clusters of correlations with visual features and semantic features (Fig. 3A-B). Nevertheless, it does not undermine the positive finding that modality-independent representations correlated with visual representations, supporting the conclusion that perceptual processes are engaged in the encoding of modality-independent concepts.

**CONCLUSION**

Our investigation into the content of modality-independent conceptual representations revealed that they contain not only semantic features, but also visual representations. Thus they are not strictly amodal, consistent with theories in which perceptual processes play a fundamental role in the encoding of modality-independent representations. We found no evidence that lexical representations contribute to the content of modality-independent representations. This suggests that activating conceptual knowledge does not necessarily involve the activation of lexical labels.

**METHODS**

**Participants**

Twenty-five native English speakers were paid to participate in the study (15 female, Age: $M$ = 23.70 SD = 4.92). Participants were recruited by posting advertisements about the study throughout the Washington Square campus of New York University (NYU), although recruitment was not limited to NYU students. All participants reported normal or corrected-to-normal vision and no history of neurological or language disorders. The study received ethical approval from the institutional review board at NYU.

**Experimental design**

Participants were asked to perform a binary animacy judgment while their brain activity was recorded using magnetoencephalography. The stimuli consisted of 16 exemplars that were presented as pictures and as words, with each exemplar repeated 40 times for each modality resulting in a total of 1280 trials. Each trial started with a fixation cross that appeared on screen for 300ms, followed by a blank screen for 300ms and finally the target picture or word appeared on screen until a response was given using a button box. All responses were given using the left index for animate exemplars and left middle finger for inanimate exemplars. Participants were instructed to give their responses as fast and as accurately as they could. The interstimulus intervals were randomly sampled from a uniform distribution with a range of 200-700ms. Stimuli were presented using Psychopy 2020.1.2 [38]. All stimuli were matched between animate and inanimate words for length, lexical frequency, number of morphemes, number of phonemes, number of phonographic neighbors, number of orthographic neighbors, number of phonological neighbors, and average bigram count [22].

**MEG acquisition and preprocessing**

Continuous MEG was recorded with a 157-channel axial gradiometer system (Kanazawa Institute of Technology) at a sampling rate of 1000 Hz with an online band-pass filter of 0.1–200 Hz. The raw data was noise-reduced with the continuously adjusted least-squares method [39] using the MEG Laboratory software 2.004A (Yokogawa Electric and Eagle Technology Corp., Japan). The data was low-pass filtered offline at 40 Hz and bad channels were identified after visual inspection, and the data for those channels were estimated using interpolation [40]. An independent component analysis was then fitted to the data using the "fastica" method, selecting components by 95 cumulative percentage of explained variance. Components related to eye-blinks, saccades, and heartbeats were then rejected manually. Epochs from −100 to 600ms from target onset were extracted and baseline correction was done using the 100ms preceding target onset. Time-locking of the epochs to the MEG triggers was adjusted using a photodiode. Evoked responses were created by averaging every 5 random repeats of each exemplar resulting in a total of 128 averaged epochs per modality and per participant. The resulting evoked responses were down sampled by averaging non-overlapping bins of 5ms.

**Analyses**

*Cross-condition Decoding*

For each of the picture and word modalities and at each timepoint, a unique feedforward neural network classifier was used to discriminate MEG response-patterns associated with each of the 16 unique basic-level concepts from the 157 sensors of the MEG data. The data was first scaled so the mean activity at each sensor was 0 with a standard deviation of 1. This was done independently at each time point. The network architecture consisted of an input layer with input size of 157, corresponding to the MEG channels, followed by three hidden layers containing 100, 50, and 10 neurons, respectively. The output layer size was 16, corresponding to the 16 unique exemplars. Rectified Linear Unit (ReLU) activation function was employed for all hidden layers. The optimization of the neural network was facilitated by the Adam solver [41]. The learning rate was set to a constant value of 0.001. Regularization was implemented through an L2 regularization term (alpha) set to 0.0001. The training progress was evaluated using accuracy as the scoring metric, aiming to maximize classification accuracy using cross-entropy loss, with 10% of the training data being reserved as a validation set. In selecting the neural network architecture, empirical validation was employed through an iterative process involving experimentation with multiple configurations. Final accuracy scores were obtained at each time point using a 5-fold cross validation. This procedure was done separately for each subject, and average accuracy scores across subjects are reported at each time point and the results are plotted in Fig. 2A-B This first step served as a sanity check that the 16-way classification returned above-chance accuracy scores within each condition. In order to investigate whether and when modality-independent representations of semantic categories occur, we assessed the extent to which classifiers trained on one modality could generalize when tested on the other modality. This was done for all pairs of timepoints, for example, the classifier trained at 100ms on the words MEG data was tested on all timepoints from -100ms to 600ms on the pictures data, following the condition generalization approach from King and Dehaene [18]. This procedure was done separately for each subject.

*Extracting brain-based modality independent representations of concepts*

The cross-condition decoding allowed us to assess the pairs of time timepoints ($t_{train}$, $t_{test}$) where a classifier trained on one modality generalized to the other modality. In order for a classifier to be modality-independent, it must have learned to extract from the MEG data representations that allows it to perform above-chance classification regardless of the input modality. The primary goal of this study was to assess the nature of modality-independent representations. Thus, by examining the latent representations captured in the last hidden layer of modality-independent classifiers, we aimed to gain insight into the nature of those representations. For each classifier trained at time $t_{train}$, if the classifier generalized at time $t_{test}$, we extracted the last hidden layer of that classifier when the test data was fed-forward into it. We then averaged those representations across repeated exemplars, resulting in a single vector embedding for each exemplar, and at each pair of timepoints ($t_{train}$, $t_{test}$) where the classifier generalized.

*Representational Similarity Analysis (RSA)*

RSA was used to investigate the content of the latent representations that were learned by the modality-independent classifiers. The RSA involved computing a Peason's correlation between representational dissimilarity matrices (RDMs) capturing pairwise cosine distances of the modality-independent representations, and hypothesis RDMs. These hypothesis RDMs, in turn, depict the pairwise distances of exemplars as predicted by a hypothesis or computational model [19]. In other words, RSA allows us to compare the structure of the stimuli under some hypothesis space to the structure of the modality-independent space captured by the classifiers. Thus, for each pair of timepoints ($t_{train}$, $t_{test}$) where the classifier generalized across conditions, we computed a modality-independent RDM and compared it to our hypotheses RDMS.

*Hypothesis RDMs*

<u>Semantic features:</u> We operationalized each exemplar as the set of features that define it, based on feature production norms by 725 participants [21]. The final hypothesis RDM was constructed by first reducing the feature space of the entire dataset from 2526 unique semantic features to 300 principal components which captured 87% of the variance, and then calculating the pairwise cosine distance of all 16 unique exemplars.

<u>Lexical frequency:</u> Our second hypothesis was that modality-independent representations include lexical representations, meaning that the lexical label of concepts is consistently activated regardless of modality. We operationalized lexical representations as the natural logarithm of the frequency of a word, as reported by the English Lexicon Project [22]. Since this measure consists of a single scalar per exemplar, the RDM was constructed using pairwise Euclidean distance. All remaining RDMs were obtained using cosine distance.

<u>Sensory features:</u> An alternative hypothesis posits that modality-independent representations capture sensory features that are consistently recruited for the representation of concepts. To test this hypothesis we operationalized the exemplars based on their visual and auditory components as measured by the experiential feature norms from [17]. This resulted in two RDMs capturing the structure of the exemplar based on their visual and auditory features.

<u>ResNet embeddings:</u> Finally, modality-independent representations that were captured by the classifier could be a reactivation of the other modality's specific visual components. For example, when participants read the word "dog", they might reactivate the mental image of the dog that was previously encountered during the experiment. While the visual features described above capture the extent to which visual representations contribute to the representation of a concept, these features are not specific to the picture stimulus of that concept that was used in the current experiment. In order to operationalize the visual representation of the specific stimuli used here, we processed the JPEG files containing pictures and word stimuli through a pre-trained convolutional neural network (ResNet) [20] and extracted for each exemplar and each modality the last hidden layer.

To ensure that each hypothesis captured a unique aspect of the representation space, we confirmed that all pairwise Pearson's correlation coefficients between hypothesis remained under 0.3.

*Group-level statistical analyses*

To evaluate classifier accuracy at the group level, for each pair of time points ($t_{train}$, $t_{test}$) a *t*-value was computed using a one-tailed one-sample t-test assessing the group-level average accuracy score against chance (0.0625). The resulting t-value map was then thresholded at a *t* value corresponding to an uncorrected *p*-value of 0.05. Clusters were formed based on direct adjacency in time (minimum timepoints in a cluster = 2), and the sum of all t-values (Σt) was computed for each resulting cluster. This procedure was then repeated by randomly permuting the data 10,000 times in order to obtain a null distribution. The Monte Carlo *p*-value was computed for each cluster in the original t map as the proportion of random permutations in which the observed Σ*t* was larger than the values from the permutation distribution. We retained clusters whose Monte Carlo *p*-value was smaller or equal to 0.05 [42]. To assess the group-level significance of the RSA, for each pair of time points ($t_{train}$, $t_{test}$) a t-value was computed using a two-tailed one-sample *t*-test assessing the group-level average Peason's correlation coefficients score against 0. A similar procedure as described above was then performed to compute a null distribution, with the exception that we retained clusters whose Monte Carlo p-value was smaller or equal to 0.05 correcting for multiple comparisons using false discovery rate [43] across all hypotheses (5).

This work was supported by the NYUAD Research Institute (Grant G1001) and The William Orr Dingwall Dissertation Fellowship in the Cognitive, Clinical and Neural Foundations of Language.

The authors declare no competing financial interests.